\documentclass[
prc
,showpacs ,amssymb,superscriptaddress,aps
]{revtex4}
\usepackage{graphicx}
\input{epsf}

\usepackage{amsmath,amssymb}
\usepackage{bm}
\newcommand{\dalm}{\kern1pt\vbox{\hrule height 0.9pt\hbox{\vrule width 0.9pt
\hskip 2.5pt\vbox{\vskip 5.5pt}\hskip 3pt\vrule width 0.3pt}\hrule height 0.3pt}
\kern1pt}

\usepackage{ulem}

\begin{document}

%\twocolumn[\hsize\textwidth\columnwidth\hsize\csname @twocolumnfals\endcsname

% For two column
%\wideabs{

\title{Effect of nuclear saturation parameters on possible maximum mass of neutron stars}

\author{Hajime Sotani}
\email{sotani@yukawa.kyoto-u.ac.jp}
\affiliation{Division of Theoretical Astronomy, National Astronomical Observatory of Japan, 2-21-1 Osawa, Mitaka, Tokyo 181-8588, Japan}
%\affiliation{Yukawa Institute for Theoretical Physics, Kyoto University, Kyoto 606-8502, Japan}

%\author{Kei Iida}
%\affiliation{Department of Natural Science, Kochi University, 2-5-1 Akebono-cho, Kochi 780-8520, Japan
%}

%\author{Kazuhiro Oyamatsu}
%\affiliation{Department of Human Informatics, Aichi Shukutoku University,
%9 Katahira, Nagakute, Aichi 480-1197, Japan
%}

\date{\today}

% Abstract
\begin{abstract}
In order to systematically examine the possible maximum mass of neutron stars, which is one of the important properties characterizing the physics in high-density region, I construct neutron star models by adopting phenomenological equations of state with various values of nuclear saturation parameters for low-density region, which are connected to the equation of state for high-density region characterized by the possible maximum sound velocity in medium. I derive an empirical formula for the possible maximum mass of neutron star. If massive neutron stars are observed, it could be possible to get a constraint on the possible maximum sound velocity for high-density region. %As a result, we find  that, comparing to the current observation of neutron star mass, the possible maximum sound velocity for high-density region should be grater than $\sim 75\%$ of speed of light in vacuum. 
\end{abstract}

%
%%%%%%%%%%%%%%%%%%%%%%%%%%%%%%%%%%%%%%%%%%%%%%%%%
%  04.40.Dg :  Relativistic stars: structure, stability, and oscillations (see also 97.60.-s Late stages of stellar evolution) 
%  21.65.Ef  :  Symmetry energy
%  26.60.Gj  :  Neutron star crust
%  21.60.-n  :   Nuclear structure models and methods
%  97.10.Sj  :   Pulsations, oscillations, and stellar seismology 
%%%%%%%%%%%%%%%%%%%%%%%%%%%%%%%%%%%%%%%%%%%%%%%%%
%]
% For two column
%}
\maketitle
%\baselineskip 24pt
%%%%%%%%%%%%%%%%%%%%%%%%%%%%%%%%%%%%%%%%%%%%%%%%
\section{Introduction}
\label{sec:I}
%%%%%%%%%%%%%%%%%%%%%%%%%%%%%%%%%%%%%%%%%%%%%%%%

Neutron stars are good candidates for investigating physics under extreme conditions. The density inside a neutron star is significantly over the nuclear saturation density, $\rho_0$. %=2.68\times 10^{14}$ g/cm$^{3}$.
This is one reason why the neutron star structure is not yet fixed, i.e., the determination of the equation of state (EOS) for neutron star matter is quite difficult (for high-density region) in the terrestrial nuclear experiments. Therefore, the observations of a neutron star itself and/or of the phenomena associated with neutron stars could provide opportunities for obtaining a constraint on the EOS and/or an imprint of the physics for high-density region.
So far, there are several attempts to extrapolate the EOS for neutron star matter from the density around saturation to much higher densities (e.g., \cite{L2012,SLB2016}). Moreover, the discoveries of $2M_\odot$ neutron stars \cite{D2010,A2013}, where  some of soft EOSs have been ruled out,  impacted the field. In particular, the appearance of hyperons in high-density region might be a crucial problem, because EOSs with hyperons are generally soft and difficult to support $2M_\odot$.

The possible maximum mass of neutron stars is one of the important properties characterizing the physics in higher density region. In Ref. \cite{H1978}, the maximum mass of neutron stars is derived as $M \simeq 6.8M_\odot$, preparing the EOS constructed in such a way that the stiffest EOS allowed from the causality for high-density region, i.e., $d p/d \rho=1$, is connected to the EOS given for lower density region at an appropriate transition density. Here, $p$ and $\rho$ are pressure and energy density (not a number density), and the transition density is adopted around $10^{14}$ g/cm$^3$. It is also discussed that the maximum mass could become larger with the EOS which is softer for low-density region and stiffer for high-density region \cite{KSF1997}. So, since the stiffness of EOS is associated with the sound velocity via $v_s^2=d p/d \rho$, the maximum mass should depend strongly on the possible maximum sound velocity, $v_s^{\rm max}$, for high-density region. With respect to the value of $v_s^{\rm max}$, the theoretical maximum value must be 1, which comes from the causality, while there is also a conjecture that $v_s^{\rm max}$ should be less than $1/\sqrt{3}$ \cite{BS2015}. Although this conjecture may be still uncertain, it could be better to consider in the range of $1/\sqrt{3}\le v_s^{\rm max}\le 1$, for discussing the dependence of possible maximum mass on $v_s^{\rm max}$. Eventually, the exact value of $v_s^{\rm max}$ would be constrained from the observations and/or theoretical arguments.

On the other hand, the properties for lower density region are relatively well constrained from the terrestrial nuclear experiments. In particular, the nuclear saturation parameters are important for expressing lower density region. It is practically known that the neutron star constructed with the central density lower than $2\rho_0$ can be described nicely with parameters constructed as the combination of the incompressibility of the symmetric nuclear matter, $K_0$, and the so-called slope parameter of nuclear symmetry energy, $L$, via $\eta=(K_0L^2)^{1/3}$ \cite{SIOO2014,SSB2016}. Therefore, it is expected that the maximum mass of neutron stars should also depend on $\eta$, where $\eta$ has been already constrained in some range via the constraints on $K_0$ and $L$ \cite{Tsang2012,Newton2014,Khan2013} and will be further constrained in the future.

Now, it is considered that the possible maximum mass of neutron stars must depend on the two parameters, i.e., $v_s^{\rm max}$ and $\eta$. In order to see such a dependence, I systematically examine the neutron star models by adopting the phenomenological EOS for lower density region with various values of $\eta$, which are connected to the EOS for higher density region characterized by $v_s^{\rm max}$. I remark that the possible maximum mass must depend on the transition density where the EOSs for lower and higher density regions are connected. According to the result in Ref. \cite{H1978}, the possible maximum mass can be inversely proportional to the square root of the transition density. Meanwhile, the properties of nuclear matter for $\rho\lesssim 2\rho_0$ are relatively known experimentally and predicted well theoretically. In fact, one could expect that the non-nucleonic components do not appear below $\sim 2\rho_0$, and that the uncertainties from three-nucleon interactions in EOS for pure neutron matter do not become significant below $\sim 2\rho_0$ \cite{GCR2012}. Thus, since the EOS above $\sim 2\rho_0$ is more uncertain, I adopt the transition density to be $2\rho_0$ as in Ref. \cite{BS2015}.
%Then, as shown in text, we will succeed to derive the empirical formula expressing the possible maximum mass of neutron stars as a function of $\eta$ and $v_s^{\rm max}$. Furthermore, comparing it to the current observation of neutron star mass, we can get the constraint on the value of $v_s^{\rm max}$, i.e., $v_s^{\rm max}$ should be grater than $\sim75\%$ of speed of light in vacuum. This result shows that one may need to prepare some mechanism with which the EOS for higher density region makes stiff, for example introducing a vector interaction.
Hereafter I adopt the units of $c = 1$, where $c$ denotes the speed of light.
%, and the metric signature is $(-, +, +, +)$ in this paper.

%%%%%%%%%%%%%%%%%%%%%%%%%%%%%%%%%%%%%%%%%%%%%%%%
\section{EOS and saturation parameters}
\label{sec:II}
%%%%%%%%%%%%%%%%%%%%%%%%%%%%%%%%%%%%%%%%%%%%%%%%

For any EOSs, the bulk energy per nucleon of uniform nuclear matter at zero temperature can be expanded around the saturation point of symmetric nuclear matter, for which the number of proton is equal to that of neutron, as a function of the baryon number density $n_{\rm b}$ and neutron excess $\alpha$, as discussed in Ref. \cite{L1981}:
\begin{equation}
  w = w_0  + \frac{K_0}{18n_0^2}(n_{\rm b}-n_0)^2 + \left[S_0 + \frac{L}{3n_0}(n_{\rm b}-n_0)\right]\alpha^2,
  \label{eq:w}
\end{equation}
where $w_0$ and $K_0$ are the saturation energy and incompressibility at the saturation density, $n_0$, of symmetric nuclear matter, while $S_0$ and $L$ are associated with the density dependent nuclear symmetry energy. $w_0$, $n_0$, and $S_0$, which are absolute values at the saturation point, are relatively constrained well via terrestrial nuclear experiments, owing to the nuclear saturation. Meanwhile, since $K_0$ and $L$ change rapidly at the saturation point, one has to obtain experimental data in a wide range of densities around the saturation point. Thus, it is more difficult to fix the values of $K_0$ and $L$ via the terrestrial experiments. For this reason, I focus on $K_0$ and $L$ as parameters characterizing EOS.

In practice, to systematically analyze the dependence of neutron star properties on the saturation parameters $K_0$ and $L$, I adopt the phenomenological EOS proposed by Oyamatsu and Iida \cite{OI2003,OI2007}. This EOS is constructed in such a way that the energy of uniform nuclear matter reproduces to the form as Eq. (\ref{eq:w}) in the limit of $n_{\rm b}\to n_0$ and $\alpha\to 0$ for various values of $y\equiv -K_0S_0/(3n_0L)$ and $K_0$. For given $K_0$ and $y$, the other saturation parameters $n_0$, $w_0$, and $S_0$ are determined to fit the empirical data for masses and radii of stable nuclei \cite{OI2003,OI2007}. Hereafter, I call this phenomenological EOS as OI-EOS. In particular, I focus on the parameter $K_0$, $L$, and $y$ in the range of $180\le K_0 \le 360$ MeV, $0<L<160$ MeV, and $y<-200$ MeV fm$^3$, which can reproduce the mass and radius data for stable nuclei well and effectively cover even extreme cases \cite{OI2003}. The concrete parameter sets adopted in this paper are shown in Table \ref{tab:EOS}, where $\eta$ is an auxiliary parameter defined as $\eta = (K_0 L^2)^{1/3}$ \cite{SIOO2014}. I remark that the low-mass neutron star models where central density is up to $\rho_c=2\rho_0$ can be described well with the parameter $\eta$ independently of the nuclear theoretical models \cite{SIOO2014,SSB2016}.

%%%%%%%%%%%%%%%%%%%%%%%%%%%%%%%%%%%
% Table 1
%%%%%%%%%%%%%%%%%%%%%%%%%%%%%%%%%%%
\begin{table}
\centering
\caption{Saturation parameters in OI-EOS and an auxiliary parameter $\eta\equiv(K_0L^2)^{1/3}$.
}
\begin{tabular}{cccccc}
\hline\hline
 & $y$ (MeV fm$^3$) & $K_0$ (MeV) & $L$ (MeV) & $\eta$ (MeV) & \\
\hline
 &   $-220$ & 180 & 52.2 & 78.9 &  \\
 &   $-220$ & 230 & 73.4 & 107.4 &  \\
 &   $-220$ & 280 & 97.5 & 138.6 &  \\
 &   $-220$ & 360 & 146.1 & 197.3 & \\
\hline
 &   $-350$ & 180 & 31.0 & 55.7  & \\
 &   $-350$ & 230 & 42.6 & 74.7  & \\
 &   $-350$ & 280 & 54.9 & 94.5  & \\
 &  $ -350$ & 360 & 76.4 & 128.1  & \\
\hline
 &   $-600$ & 230 & 23.7 &  50.6 & \\
 &   $-600$ & 280 & 30.1 & 63.4  & \\
 &  $ -600$ & 360 & 40.9 & 84.4  & \\
\hline\hline
\end{tabular}
\label{tab:EOS}
\end{table}
%%%%%%%%%%%%%%%%%%%%%%%%%%%%%%%%%%%

On the other hand, several EOSs have been suggested for the density region higher than $\sim 2\rho_0$, which are based on the different nuclear theories, interactions, and components. The theoretical constraints on EOS are only that the sound speed should be less than the speed of light (causality), and that the sound speed should be more than zero (thermodynamics stability). So, the stiffest EOS satisfying the theoretical constraints can be expressed in the density region of $\rho>\rho_t$, such as 
\begin{equation}
   p = \rho - \rho_t + p_t,  \label{eq:eos1}
\end{equation}
where $\rho_t$ is a transition density and $p_t$ is the pressure determined at $\rho=\rho_t$ with the EOS for lower density region. Adopting this type of EOS for high-density region and connecting to the Harrison-Wheeler EOS for lower density at $\rho_t=4.6\times 10^{14}$ g/cm$^3$, the maximum mass of neutron star is expected as $M_{\rm max}\simeq 3.2 M_\odot$ \cite{H1978,ST1983}. However, since the stellar properties in the density region of $\rho\lesssim 2\rho_0$ strongly depend on $\eta$ \cite{SIOO2014,SSB2016}, the maximum mass of the neutron star could also depend on $\eta$ even if one fixes the transition density $\rho_t$.

In addition, it has been conjectured that the sound velocity inside the star should be smaller than the speed of light in vacuum divided by $\sqrt{3}$ \cite{BS2015}. With this conjecture, the stiffest EOS for higher density region can be expressed as $p = (\rho - \rho_t)/3 + p_t$. Since this EOS becomes softer than the EOS given by Eq. (\ref{eq:eos1}), the possible maximum mass becomes smaller. In practice, the neutron star mass was discussed with this conjecture for $\rho_t=2\rho_0$, where the possible maximum mass is $\sim 2M_\odot$ \cite{BS2015}. So, if the neutron star more massive than $\sim 2M_\odot$ were to be discovered, this conjecture may not be good. In fact, a candidate of massive neutron star has been discovered in a neutron star and white dwarf binary system, where the mass of neutron star is estimated as $M=(2.74 \pm 0.21) M_\odot$ \cite{Freire2008}.

In any way, the possible maximum sound velocity inside the star, which is associated with the stiffness of EOS, must affect the determination of maximum mass of neutron stars. Thus, in order to examine the dependence of the possible maximum sound velocity inside the star (or the stiffness of EOS) on the maximum mass of neutron star, I consider a general formula of EOS given by
\begin{equation}
   p = \alpha(\rho - \rho_t) + p_t,  \label{eq:eos2}
\end{equation}
where $\alpha$ is an parameter associated with the possible maximum sound velocity inside the star, i.e., $v_s^{\rm max}=\sqrt{\alpha}$ \cite{ST1983}. For this examination, I adopt the OI-EOS for lower density region up to $\rho_t= 2\rho_0$, i.e., $p_t$ is the pressure determined with OI-EOS at the transition density $\rho_t$, and I adopt the EOS given by Eq. (\ref{eq:eos2}) for $\rho>2\rho_0$. I remark that I simply connect the EOS for lower and higher density regions at the transition density as in Ref. \cite{BS2015}. Thus, the EOS is almost continuous, but the sound velocity is not continuous at the transition density. In this paper, I focus on $\alpha$ in the range of $1/3\le \alpha\le 1$. Then, I will see the dependence of the maximum mass on $\eta$ and $\alpha$.

%%%%%%%%%%%%%%%%%%%%%%%%%%%%%%%%%%%%%%%%%%%%%%%%
\section{Possible maximum mass}
\label{sec:III}
%%%%%%%%%%%%%%%%%%%%%%%%%%%%%%%%%%%%%%%%%%%%%%%%

\begin{figure*}
\begin{center}
\includegraphics[scale=0.53]{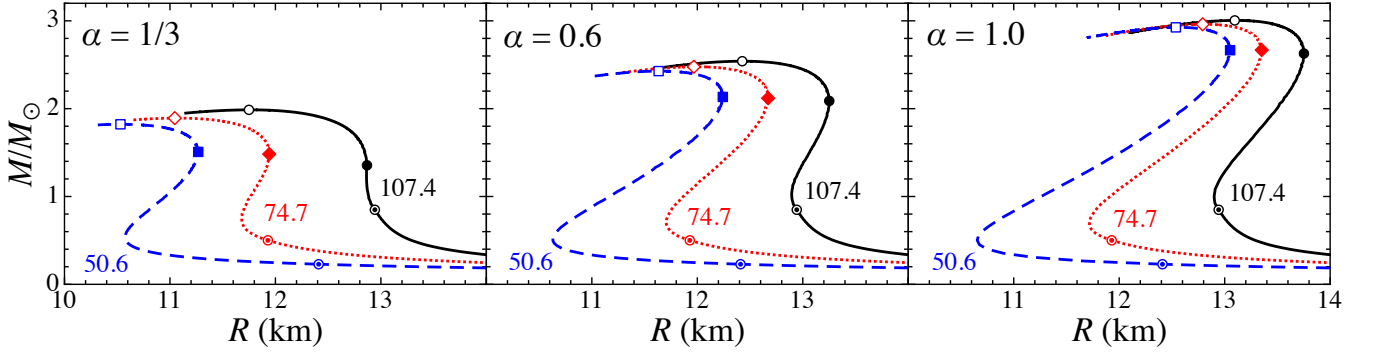} 
\end{center}
\caption{%%
Mass and radius relation for various EOSs for lower density region with $\eta=50.6$ (dashed line), 74.7 (dotted line), and 107.4 (solid line). The left, middle, and right panels correspond to different sound velocities for higher density region, i.e., $\alpha=1/3$, 0.6, and 1, respectively. The open marks correspond to the stellar models with maximum mass for various EOS models, while the filled marks correspond to the stellar models with local maximum radius. For reference, the stellar models constructed with the central density $\rho_c=2\rho_0$ denote by the double circles. 
}
\label{fig:MR}
\end{figure*}

The spherically symmetric neutron star models are constructed by integrating the Tolman-Oppenheimer-Volkoff equation together with the appropriate EOS. As an example, in Fig. \ref{fig:MR}, I show the relation between the stellar mass and radius for the cases of $\alpha=1/3$, $0.6$, and $1$ with $\eta=50.6$, $74.7$, and $107.4$, 
%Each line in the figure corresponds to the stellar model constructed with the central density, which is in the region of $2\rho_0\le\rho_c\le 3\times 10^{15}$ g/cm$^3$. That is, the rightmost edge of each line corresponds to the stellar model constructed with $\rho_c=2\rho_0$ adopting the OI-EOS with various values of $\eta$. 
where open marks denote the maximum masses for various EOS models. From this figure, I find that the maximum mass strongly depends on the possible maximum sound velocity inside the star, while the dependence on $\eta$ is relatively weak. Additionally, the filled marks in the figure denote the local maximum of the stellar radius for various EOS models, which tells us that the local maximum radius becomes larger with $\alpha$.

%%%%%%%%%%%%%%%%%%%%%%%%%%%%%%%%%%%
% Figure 1
%%%%%%%%%%%%%%%%%%%%%%%%%%%%%%%%%%%
%%%%%%%%%%%%%%%%%%%%%%%%%%%%%%%%%%%

\begin{figure}
\begin{center}
\includegraphics[scale=0.53]{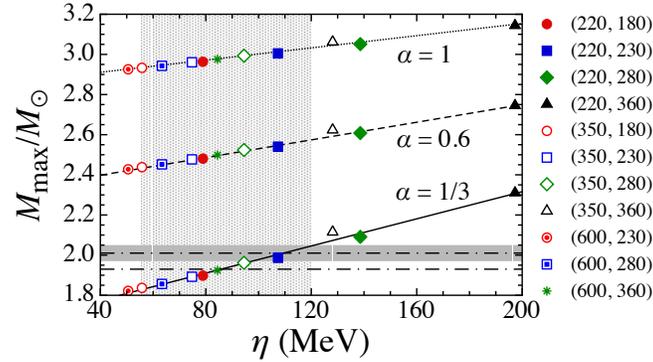} 
\end{center}
\caption{%%
The expected maximum masses for various EOS models are shown with different marks, while the solid, dashed, and dotted lines respectively denote the fitting formula given by Eq. (\ref{eq:Meta}) for $\alpha=1/3$, 0.6, and 1. In the label, I show the values of the saturation parameters for the adopted EOS, such as $(-y,K_0)$. The region between the horizontal dot-dash-lines denotes the mass observation of PSR J1614-2230 \cite{D2010}, while the horizontal shaded region denotes the mass observation of PSR J0348+0432 \cite{A2013}. The stippled region denotes a plausible range for $\eta$ determined from the current terrestrial nuclear experiments.
}
\label{fig:etaM}
\end{figure}

To see the dependence of the maximum mass on $\eta$, in Fig. \ref{fig:etaM} I plot the maximum mass predicted from the various values of $\eta$ for the cases of $\alpha=1/3$, $0.6$, and $1$. From this figure, one can observe that the maximum mass with fixed value of $\alpha$ is well fitted as a linear function of $\eta$, such as
\begin{equation}
  \frac{M_{\rm max}}{M_\odot} = a_1 + a_2  \left(\frac{\eta}{1\ {\rm MeV}}\right), \label{eq:Meta}
\end{equation}
where $a_1$ and $a_2$ are coefficients in the linear fitting, depending on the value of $\alpha$. In Fig. \ref{fig:etaM}, the linear fitting given as Eq. (\ref{eq:Meta}) for $\alpha=1/3$, $0.6$, and $1$ are shown with the solid, dashed, and dotted lines, respectively. %We also find that the dependence of the maximum mass on $\eta$ becomes stronger for smaller value of $\alpha$.

With respect to the value of $\eta$, by adopting fiducial values of $30\lesssim L \lesssim 80$ MeV \cite{Newton2014} and $K_0=230\pm 40$ MeV \cite{Khan2013}, one can get a plausible range for $\eta$ as $55.5\lesssim \eta \lesssim 120$ MeV. This plausible range of $\eta$ is shown in Fig. \ref{fig:etaM} as the stippled region, while the observations of neutron star masses, i.e., $M=(1.97\pm 0.04) M_\odot$ \cite{D2010} and $M=(2.01\pm 0.04) M_\odot$ \cite{A2013}, are also shown in the same figure. To explain the observations of neutron star masses, the case with $\alpha=1/3$, which comes from the conjecture of Ref. \cite{BS2015}, seems to be marginal with the plausible range of $\eta$. In practice, in order to explain the lower limit of neutron star mass of PSR J0348+0432, i.e., $M=1.97M_\odot$, $\eta$ should be larger than $\sim 100$ MeV, which leads to the constraint of $L\gtrsim 66$ MeV with adopting the canonical value of $K_0=230$ MeV.

%%%%%%%%%%%%%%%%%%%%%%%%%%%%%%%%%%%
% Figure 2
%%%%%%%%%%%%%%%%%%%%%%%%%%%%%%%%%%%
%%%%%%%%%%%%%%%%%%%%%%%%%%%%%%%%%%%

\begin{figure}
\begin{center}
\includegraphics[scale=0.53]{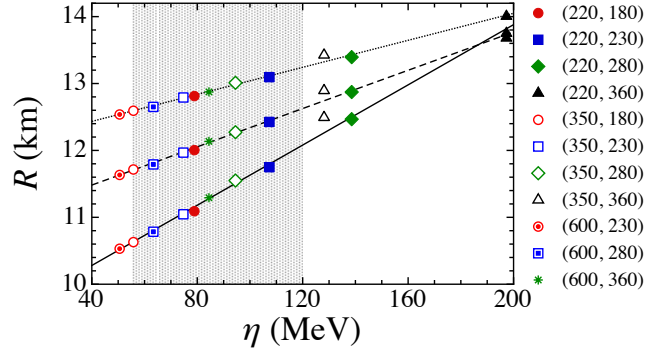} 
\end{center}
\caption{%%
The expected radii for the stellar models with maximum mass constructed with various EOS models are shown with different marks, while the solid,  dashed, and dotted lines respectively denote the fitting formula given by Eq. (\ref{eq:Reta}) for $\alpha=1/3$, 0.6, and 1. In the label, I show the values of the saturation parameters for the adopted EOS, such as $(-y,K_0)$. The stippled region denotes a plausible range for $\eta$ determined from the current terrestrial nuclear experiments.
}
\label{fig:etaR}
\end{figure}

In the similar way to the discussion about the maximum mass, I additionally find that the radius for the neutron star with maximum mass with fixed $\alpha$ can be well described as a linear function of $\eta$ as shown in Fig. \ref{fig:etaR}. Thus, I can get a linear fit, such as
\begin{equation}
  \frac{R}{1\ {\rm km}} = b_1 + b_2  \left(\frac{\eta}{1\ {\rm MeV}}\right),  \label{eq:Reta}
\end{equation}
where $b_1$ and $b_2$ are coefficients in the linear fit, which depend on the value of $\alpha$.

%%%%%%%%%%%%%%%%%%%%%%%%%%%%%%%%%%%
% Figure 3
%%%%%%%%%%%%%%%%%%%%%%%%%%%%%%%%%%%
%%%%%%%%%%%%%%%%%%%%%%%%%%%%%%%%%%%

\begin{figure}
\begin{center}
\includegraphics[scale=0.53]{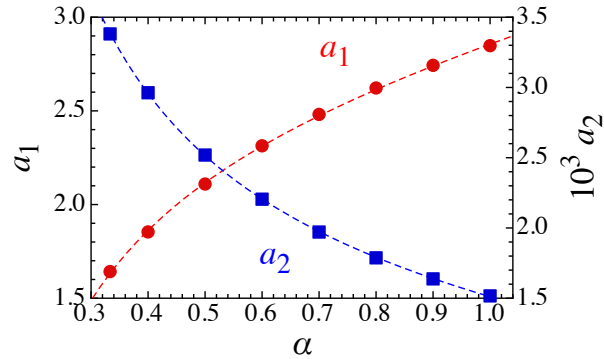} 
\end{center}
\caption{%%
The coefficients $a_1$ and $a_2$ in Eq. (\ref{eq:Meta}) as a function of $\alpha$. The dashed lines are fitting given by Eqs. (\ref{eq:a1a}) and (\ref{eq:a2a}).
}
\label{fig:a1a2}
\end{figure}

\begin{figure}
\begin{center}
\includegraphics[scale=0.53]{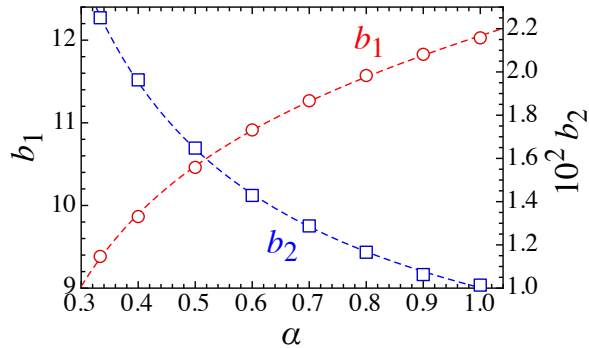} 
\end{center}
\caption{%%
The coefficients $b_1$ and $b_2$ in Eq. (\ref{eq:Reta}) as a function of $\alpha$. The dashed lines are fitting given by Eqs. (\ref{eq:b1a}) and (\ref{eq:b2a}).
}
\label{fig:b1b2}
\end{figure}

Furthermore, I plot the coefficients in the linear fit [Eqs. (\ref{eq:Meta}) and (\ref{eq:Reta})], i.e., $a_1$, $a_2$, $b_1$, and $b_2$, as a function of $\alpha$ in Figs. \ref{fig:a1a2} and \ref{fig:b1b2}. Then, I find that such coefficients can be well fitted as a function of $\alpha$ with the functional forms given by
\begin{eqnarray}
   a_1(\alpha) &=& -0.356 / \alpha + 2.445 + 0.767 \alpha, \label{eq:a1a} \\ 
   a_2(\alpha) &=& \left(0.806 / \alpha + 1.098 -0.393 \alpha\right) \times 10^{-3},  \label{eq:a2a} \\
   b_1(\alpha) &=& -0.883 / \alpha + 11.548 + 1.388 \alpha, \label{eq:b1a} \\ 
   b_2(\alpha) &=& \left(6.008 / \alpha + 4.834 -0.824 \alpha\right) \times 10^{-3}. \label{eq:b2a}    
\end{eqnarray}
In Figs. \ref{fig:a1a2} and \ref{fig:b1b2}, the marks denote the numerical values in linear fitting [Eqs. (\ref{eq:Meta}) and (\ref{eq:Reta})], while the dashed lines are plotted with using Eqs. (\ref{eq:a1a}) -- (\ref{eq:b2a}). Now, I can get the fitting formulae expressing the maximum mass and radius of neutron star with maximum mass as a function of $\eta$ and $\alpha$.

%%%%%%%%%%%%%%%%%%%%%%%%%%%%%%%%%%%
% Figure 4
%%%%%%%%%%%%%%%%%%%%%%%%%%%%%%%%%%%
%%%%%%%%%%%%%%%%%%%%%%%%%%%%%%%%%%%

%%%%%%%%%%%%%%%%%%%%%%%%%%%%%%%%%%%
% Figure 5
%%%%%%%%%%%%%%%%%%%%%%%%%%%%%%%%%%%
%%%%%%%%%%%%%%%%%%%%%%%%%%%%%%%%%%%

\begin{figure}
\begin{center}
\includegraphics[scale=0.53]{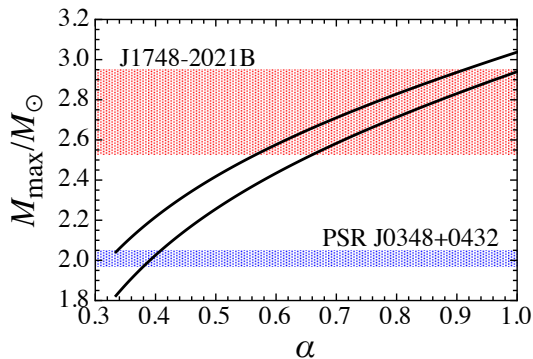} 
\end{center}
\caption{%%
The maximum mass of neutron star predicted with the plausible value of $\eta$, i.e., $55.5\lesssim \eta\lesssim 120$ MeV, is shown as a function of $\alpha$ in the region between the solid lines. The shaded regions correspond to the observations of neutron star mass for J1748$-$2021B \cite{Freire2008} and for PSR J0348+0432 \cite{A2013}.
}
\label{fig:alpha-M}
\end{figure}

Finally, adopting the linear fitting expressed by Eq. (\ref{eq:Meta}) together with Eqs. (\ref{eq:a1a}) and (\ref{eq:a2a}), one can obtain the possible maximum mass of neutron star predicted with the plausible value of $\eta$ for various values of $\alpha$, which corresponds the region between the solid lines in Fig. \ref{fig:alpha-M}. In the same figure, I put the observations of neutron star mass for J1748$-$2021B \cite{Freire2008} and for PSR J0348+0432 \cite{A2013} with the shaded regions. As mentioned the before, the observation of PSR J0348+0432 is possible to explain even for $\alpha = 1/3$ (or $v_s^{\rm max}=1/\sqrt{3}$), but to explain the observation of J1748$-$2021B (even though this may be rather uncertain), the value of $\alpha$ should be at least lager than $\sim 0.57$, i.e., $v_s^{\rm max}\gtrsim 0.75c$. If this result is to be believed, one may need to introduce some mechanism with which the EOS for higher density region makes stiff, for example introducing a vector interaction. In any way, with future observations of massive neutron stars, one could put a constraint on the possible maximum sound velocity inside a star.

%%%%%%%%%%%%%%%%%%%%%%%%%%%%%%%%%%%
% Figure 6
%%%%%%%%%%%%%%%%%%%%%%%%%%%%%%%%%%%
%%%%%%%%%%%%%%%%%%%%%%%%%%%%%%%%%%%

%%%%%%%%%%%%%%%%%%%%%%%%%%%%%%%%%%%%%%%%%%%%%%%%
\section{Conclusion}
\label{sec:IV}
%%%%%%%%%%%%%%%%%%%%%%%%%%%%%%%%%%%%%%%%%%%%%%%%

To describe the EOS of neutron star matter, the nuclear saturation parameters are important for low-density region, while the possible maximum sound velocity could be a key parameter for high-density region. In fact, the neutron star structures in the density region lower than $\sim 2\rho_0$ can be well described by the combination of the nuclear saturation parameters such as $\eta=(K_0 L^2)^{1/3}$ \cite{SIOO2014,SSB2016}. In order to discuss the possible maximum mass of neutron stars, I simply consider the EOS constructed in such a way that the phenomenological EOS with various values of $\eta$ for lower density region is connected at $\rho=2\rho_0$ to the EOS for higher density region characterized by the possible maximum sound velocity. As a result, I find that the possible maximum mass can be expressed as a function of $\eta$ and the possible maximum sound velocity for high-density region. %The current observation of neutron star mass tells us at least that the possible maximum sound velocity inside the star should be larger than $\sim 75\%$ of speed of light in vacuum, adopting the plausible value of $\eta$. 
With future observations of massive neutron stars, one could get a constraint on the possible maximum sound velocity, which may give us a hint for understanding the physics in the higher density region. In this paper I simply connect the EOS for lower density region to that for higher density region, but the smooth connection at the transition density might reduce the maximum mass. In such a case, the constraint on the possible maximum sound velocity inside a star may become more severe. %We remark that the maximum mass should depend on the transition density $\rho_t$, but there are many uncertainties for the density region higher than $\sim 2\rho_0$. If one consider  where one needs to introduce some additional parameter to express the EOS

%\newpage
%%%%%%%%%%%%%%%%%%%%%%%%%%%%%%%%%%%%%%%%%%%%%%%%
\acknowledgments
%%%%%%%%%%%%%%%%%%%%%%%%%%%%%%%%%%%%%%%%%%%%%%%%
I am grateful to K. D. Kokkotas and J. M. Lattimer for fruitful comments, and to K. Oyamatsu and K. Iida for preparing the EOS table. I am also grateful to B. Balantekin for checking the manuscript carefully. This work was supported in part by Grant-in-Aid for Young Scientists (B) through Grant No. 26800133 provided by Japan Society for the Promotion of Science (JSPS) and by Grants-in-Aid for Scientific Research on Innovative Areas through Grant No. 15H00843 provided by MEXT.

%\appendix
%%%%%%%%%%%%%%%%%%%%%%%%%%%%%%%%%%%%%%%%%%%%%%%%
%\section{}   % Appendix A
%\label{sec:appendix_1}
%%%%%%%%%%%%%%%%%%%%%%%%%%%%%%%%%%%%%%%%%%%%%%%%

%%%%%%%%%%%%%%%%%%%%%%%%%%%%%%%%%%%%%%%%%%%%%%%%

\end{document}